\newif\ifdraft\draftfalse
\newif\ifarxiv\arxivtrue
\newif\ifcdfnote\cdfnotefalse
\newif\iffnalconf\fnalconffalse

\documentclass[slac_one]{revtex4}

\usepackage{graphicx}
\usepackage{fancyhdr}
\newcommand{\cvsstamp}{$\ $Date: 2008/10/10 04:15:17 $\ \bullet\; $RCSfile: gmetICHEP08.tex,v $\
\bullet\; $Revision: 1.7 $\ \bullet\; $Author: slava77 $\ $ }

\def\mrm{\mathrm}

\newcommand{\eV}{\ensuremath{\mathrm{e\kern -0.1em V}}}
\newcommand{\TeV}{\ensuremath{\mathrm{Te\kern -0.1em V}}}
\newcommand{\GeV}{\ensuremath{\mathrm{Ge\kern -0.1em V}}}
\newcommand{\MeV}{\ensuremath{\mathrm{Me\kern -0.1em V}}}
\newcommand{\keV}{\ensuremath{\mathrm{ke\kern -0.1em V}}}
\newcommand{\GeVc}{\ensuremath{\GeV/c}}

\newcommand{\keVcc}{\ensuremath{\keV/{c^2}}}

\newcommand{\Tesla}{\ensuremath{\mathrm{T}}}

\newcommand{\ns}{\ensuremath{\mathrm{ns}}}

\newcommand{\pbin}{\ensuremath{\mathrm{pb}^{-1}}}
\newcommand{\fbin}{\ensuremath{\mathrm{fb}^{-1}}}

\newcommand{\ppbar}{\ensuremath{p\overline{p}}}

\newcommand{\tevE}{\ensuremath{\sqrt{s} = 1.96~\TeV}}

\newcommand{\sumpt}{\ensuremath{\sum p_T}}

\newcommand{\met} {\mbox{${E\!\!\!\!/_T}$}}

\newcommand{\gmet}{\ensuremath{\gamma+\met}}
\newcommand{\gmetjet}{\ensuremath{\gmet+\mrm{jet}}}
\newcommand{\jmet}{\ensuremath{\mrm{jet}+\met}}

\begin{document}

\title{
\ifcdfnote
\vspace{-2cm}
\begin{flushright}\large{
CDF/PUB/EXOTIC/PUBLIC/9545\\
09/30/2008
}
\end{flushright}
\vspace{-0cm}
\fi
\iffnalconf
\begin{flushright}\large{
FERMILAB-CONF-XX-YYY-Z\\
\today
}
\end{flushright}
\vspace{0.25cm}
\fi
Searches for New Physics in $\gamma+\met$ Events at CDF Run~II} 
\ifdraft
\cvsstamp
\fi

%

\author{V.~Krutelyov (for the CDF Collaboration)}
\affiliation{UCSB, Santa Barbara, CA 93106, USA}

\begin{abstract}
The addition of the EMTiming system installed
 to provide the time measurements of the electromagnetic calorimeter signals
has significantly increased the sensitivity of CDF
to events with \gmet.
Here I review recent searches in this signature
performed by CDF using data from \ppbar\ collisions at \tevE.
They provide new constraints on models with  large extra dimensions
and with  gauge mediated supersymmetry breaking.
\end{abstract}

\boldmath
\maketitle
\unboldmath

\thispagestyle{fancy}


\section{INTRODUCTION} 
\label{sec:intro}
Searches for physics beyond the standard model (SM)
in hadron collisions cover a variety
of final states, in particular in events with
a photon and a large missing transverse
energy (\met)~\cite{unitDefs}.
The large \met\ is associated with particles that weakly interact with matter
and leave the detector undetected.
In supersymmetric (SUSY) models they are  the lightest supersymmetric particles (LSP),
the natural candidates to account for the dark matter~\cite{PDG}.
In models with large extra dimensions which provide
a solution to the hierarchy problem they are the Kaluza-Klein gravitons~\cite{ADD}.
A photon can be produced at the collision point (prompt photon)
or  away from it, 
like in a decay of a heavy long-lived particle,
in which case it is detected later in time (delayed photon).
Other contributions to the \gmet\ signature in CDF are from non-collision sources:
a high energy muon from beam
or from cosmic ray interactions can radiate in the detector and be misidentified as a photon (a beam halo
or a cosmic photon respectively).
Each type of photons has a unique  distribution of the time they reach
the electromagnetic calorimeter.
This time can be measured by
the EMTiming system installed in 2005~\cite{EMTnim}.

Here I present two analyses which benefit the most from the EMTiming system.
The first analysis is a search for delayed photons motivated by the model
with gauge mediated supersymmetry breaking (GMSB), performed
in the \gmetjet\ signature using 570~\pbin\ of data~\cite{llxPRL}.
The second is a search in the exclusive \gmet\ signature
motivated by the model with large
extra dimensions (LED), performed using 2~\fbin\ of data~\cite{ledCDFprl}.

\section{CDF DETECTOR}
\label{sec:cdf}
A detailed description of the CDF detector can be found in Ref.~\cite{tdrCDF}.
The vertexing and the tracking detectors surrounding the interaction region
placed in solenoidal magnetic field of 1.4~\Tesla\ 
are used to reconstruct trajectories of charged particles and measure their
momenta and points of origin.
Further out are the calorimeters with electromagnetic and hadronic
longitudinal segmentation
used to identify and measure  photons, electrons, and jets.
The EMTiming system is used to measure the calorimeter signal time
with resolution of 0.5~\ns\ and a threshold of about 3~\GeV.
Placed outside the calorimeter
are the muon detectors used to identify muons from the collisions and cosmic rays.


\section{SEARCH FOR HEAVY LONG-LIVED NEUTRALINOS}
\label{sec:gmsb}
A search for delayed photons~\cite{llxPRL} is
motivated by the GMSB model
in which the lightest neutralino $\tilde{\chi}_1^0$
decays to a photon and a gravitino $\tilde{G}$.
Given the resolution of the EMTiming system and the CDF detector geometry
the sensitivity to the delayed photons is the best
for a proper lifetime of the $\tilde{\chi}_1^0$ in the nanosecond range. 
This range is consistent with
the mass of the gravitino to be in the \keVcc\ range which would explain the dark matter abundance.
In \ppbar\ collisions the  $\tilde{\chi}_1^0$'s are produced in chain decays of heavier particles 
and the events with \gmet\  are expected to have some additional activity
in the calorimeter.

The pre-selected events are required to have an isolated photon with $E_T>30~\GeV$
and $|\eta|<1$, the missing energy of $\met>30~\GeV$, a jet with $E_T^{\mrm{jet}}>30~\GeV$ 
and $|\eta^{\mrm{jet}}|<2$, and a vertex with track-\sumpt\ above 15~\GeVc~\cite{unitDefs}.
The backgrounds are from events with prompt, beam halo, and cosmic photons.
The photon time $t_c^\gamma$, measured relative to the vertex time,
is used to discriminate the signal from the backgrounds.
The contributions from each background are predicted from the $t_c^\gamma$ distributions
extracted from data outside the region of $1.2~\ns < t_c^\gamma < 10~\ns$.
The expected contribution from the GMSB is estimated using simulation 
for points in the $(m_{\tilde{\chi}_1^0}, c\tau_{\tilde{\chi}_1^0})$ space.

\begin{figure*}[b]
\centering
\begin{tabular}{cc}
\includegraphics[width=0.44\textwidth]{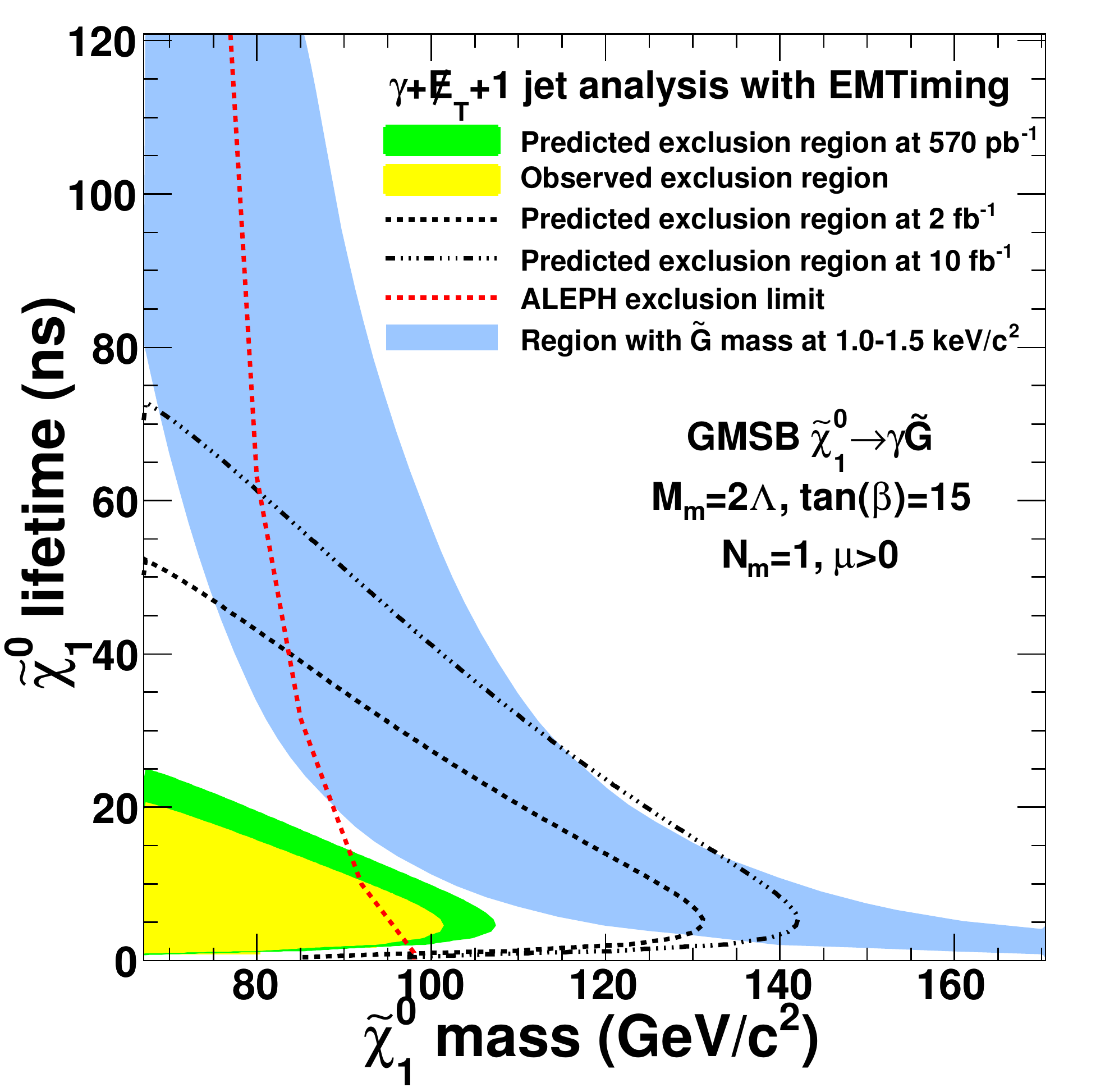}
&
\includegraphics[width=0.52\textwidth]{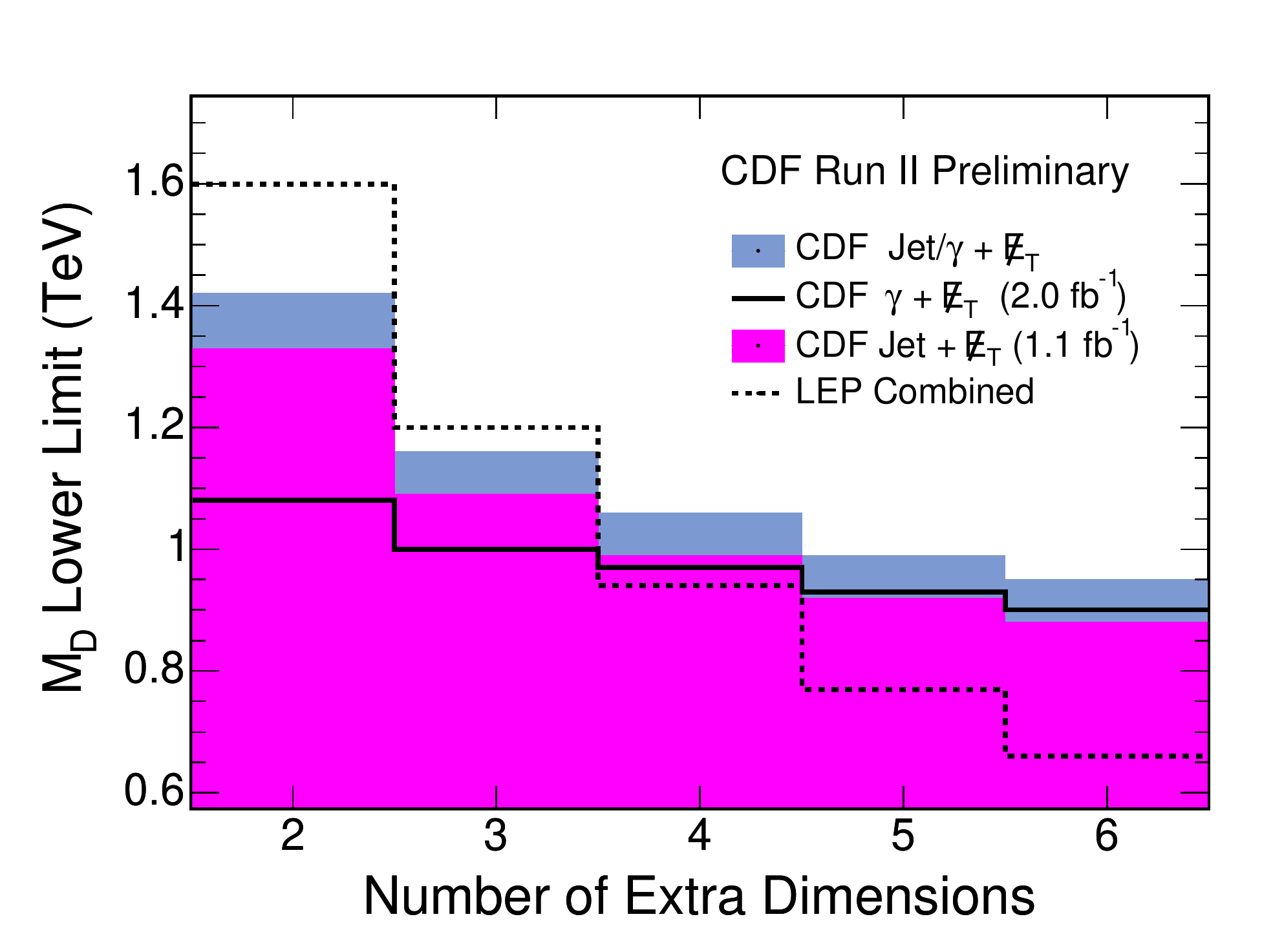}
\end{tabular}
\caption{Predicted and observed region in
$(m_{\tilde{\chi}_1^0}, c\tau_{\tilde{\chi}_1^0})$ excluded at 95\% CL
in the delayed photon analysis
together with the expected sensitivity for a future analysis
compared to results from LEP and the region favored by cosmological observations (left).
Observed 95\% CL upper limit on $M_D$ for different numbers of extra dimensions
using \gmet\ events
together with the limit obtained from the combination with a similar analysis of
\jmet\ events compared to the limit from LEP (right).
}
 \label{fig:limits}
\end{figure*}

The optimal event selection in $\met, E_T^{\mrm{jet}}, t_c^\gamma$,
and the angular separation between the \met\ and the jet vectors $\Delta\phi(\met, \mrm{jet})$
is chosen to provide the minimal expected upper limit on the $\tilde{\chi}_1^0$ production cross section
at the 95\% confidence level (CL) for the highest $m_{\tilde{\chi}_1^0}$ expected to be excluded.
It is found to be $\met>40~\GeV, E_T^\mrm{jet}>35~\GeV, \Delta\phi(\met, \mrm{jet})> 1~\mrm{rad},$ 
and $2~\ns < t_c^\gamma < 10~\ns$.
Two events pass this selection compared to $1.3\pm 0.7$ events expected
from $0.7\pm 0.6$, $0.5\pm 0.1$, and $0.1\pm 0.1$ events of the prompt, cosmic, and beam halo
photon backgrounds respectively.
The expected and observed exclusion regions are shown in Fig.~\ref{fig:limits}, together with
the exclusion region expected from a future analysis
which would probe the region where the gravitino
can be the dark matter particle.

\section{SEARCH FOR LARGE EXTRA DIMENSIONS}
\label{sec:led}
A search in events with \met\ and a prompt photon~\cite{ledCDFprl}
is motivated by the LED model~\cite{ADD}.
In this case a graviton produced in a \ppbar\ collision recoils against a photon
and leaves the detector undetected.
Since the graviton momentum can have a component in extra spacial dimensions, it is visible
in our space-time as a massive particle.
In the LED model
the graviton
mass spectrum is essentially continuous and
a large number of modes
contribute to high energy interactions with a production rate potentially observable at the Tevatron
depending on the number of extra dimensions $n$ and the fundamental mass parameter $M_D$.

Pre-selected \gmet\ events are required to have one photon with $E_T>40~\GeV$ and $|\eta|<1$
and  $\met>50~\GeV$.
Events from non-collision sources are suppressed by requiring a collision vertex with at least 3 tracks,
and the photon time to be within 3~\ns\ from the expected collision time.
Contributions from cosmic rays are additionally suppressed using
hits in the muon systems and measurements in the calorimeter.
Contributions from the beam halo are suppressed by topological event selections to a negligible level.
The remaining backgrounds  are coming from \ppbar\ collisions.
The only irreducible background coming from $Z\gamma$ production, where $Z\to\nu\bar{\nu}$, 
is estimated from simulation.
Other prompt photon backgrounds are either from events with
particles misidentified as a photon or from events with a mismeasured
\met\ from an object (a lepton, photon, or a jet) lost in an uninstrumented region.
These backgrounds are suppressed by
requiring
no jets with $E_T>15~\GeV$ and no tracks with $p_T>10~\GeVc$.
The contribution from $W\to e\nu$, where the electron is identified as 
the photon, is estimated from the electron sample
scaled by the misidentification rate.
The contribution from cosmics is estimated using events with the photon time
away from the collision time.
Except for $W\to\tau\nu$, which is estimated from simulation,
the number of events with a lost object is derived from events in data with
this object identified, scaled by the simulated rate for it to be lost.

After a selection optimized for the best 95\% upper limit,
corresponding to  the photon $E_T>90~\GeV$,
40 events are observed compared to $46.3\pm 3.0$ expected from the background contributions.
The expected number of background events are 
$ 24.8 \pm 2.8$, $ 9.8 \pm 1.3$, $7.3 \pm 1.5$, and $3.6 \pm 0.4 $ coming from
$Z\gamma\to\nu\bar\nu\gamma$, cosmics, a lost object, and a lepton from a $W$ decay 
misidentified as a photon respectively.
The constraints on the LED shown in Fig.~\ref{fig:limits}
are comparable to those provided by a similar analysis of \jmet\ events
and can be combined for a better limit.

\section{SUMMARY}
\label{sec:summary}
Among the recent searches for physics beyond the SM in events with \gmet,
the search for delayed photons motivated by the GMSB and the search for prompt photons motivated by 
the LED
benefit the most from the measurements provided by the EMTiming system
installed at CDF in 2005.
The search for delayed photons employs purely data-driven background estimation
methods and provides the best reach in $\tilde{\chi}_1^0$ mass
with a proper lifetime of about 5~\ns.
Similar analysis performed with more data is expected to probe extensively the region
consistent with the gravitino being the dark matter particle.
The search in the exclusive \gmet\ signature provides constraints on the LED model
comparable to those from a similar analysis in the \jmet\ signature.

\begin{acknowledgments}
I would like to acknowledge the funding institutions supporting
the CDF Collaboration.
The full list of agencies can be found in, e.g,~\cite{llxPRL}.
\end{acknowledgments}


\begin{thebibliography}{99}
\bibitem{unitDefs}
The polar coordinate system is used with 
the $z$-axis along the proton beam.
The pseudorapidity is defined as $\eta=-\ln[\tan(\theta/2)]$.
The transverse momentum is defined as $p_T=p\cdot\sin{\theta}$,
where $p$ is the momentum. 
The transverse energy is defined as $E_T = E\cdot\sin{\theta}$, 
where $E$ is the energy measured by the calorimeter.
The missing $E_T$ is defined as $\vec{\met} = - \sum_i E_T^i \hat{n}_i$,
where the sum is taken over calorimeter cells,
and $\hat{n}_i$ is a unit vector in $xy$-plane
pointing to the cell from the origin.
\end{thebibliography}
\end{document}